\documentstyle[epsf]{l-aa}
\begin{document}
\thesaurus{06(06.13.1; 02.03.1; 02.13.1)}
\title{Dynamos with different formulations of a dynamic $\alpha$--effect}
\author{Eurico Covas\thanks{e-mail: E.O.Covas@qmw.ac.uk}\inst{1}
\and    Andrew Tworkowski\thanks{e-mail: A.S.Tworkowski@qmw.ac.uk}\inst{1}
\and    Axel Brandenburg\thanks{e-mail: Axel.Brandenburg@ncl.ac.uk}\inst{2}
\and    Reza Tavakol\thanks{e-mail: reza@maths.qmw.ac.uk}\inst{1}}
\institute{Astronomy Unit, School of Mathematical Sciences,
Queen Mary and Westfield College, Mile End Road, London E1 4NS, UK
\and Department of Mathematics and Statistics, University of Newcastle
upon Tyne NE1 7RU, UK}
\date{Received April 16; accepted June 4, 1996}
\offprints{Eurico Covas}
\maketitle 
\begin{abstract}
We investigate the behaviour of $\alpha\Omega$ dynamos with a dynamic
$\alpha$, whose evolution is governed by the imbalance between a driving
and a damping term. We focus on truncated 
versions of such dynamo models 
which are often studied in connection with
solar and stellar variability.
Given the approximate nature of such models, it is important to study how
robust they are with respect to reasonable changes in the formulation of
the driving and damping terms. For each case, we also study the effects of
changes of the dynamo number and its sign, the truncation order and
initial conditions.
Our results show that changes in the formulation of the driving term have
important consequences for the dynamical behaviour of such
systems, with the detailed nature of these effects depending
crucially on the form of the driving term assumed,
the value and the sign of the dynamo number and the 
initial conditions. On the other hand, the change in the damping term
considered here seems to produce little qualitative effect.
\end{abstract}
\section{Introduction}
It is commonly believed that the observed solar and
stellar variabilities have their origin in the hydromagnetic dynamos
associated with turbulent convection zones. Numerical studies have
been made using the full magneto-hydrodynamical partial differential 
equations (PDE), which reproduce some features
of solar and stellar dynamos (e.g. Gilman 1983).
Such models are fairly complex and do not allow extensive parameter surveys.
As a result, a number of alternatives to the 
direct integration of PDE have been pursued. 
Among these has been the employment of the mean field dynamo 
formalism (Krause \& R\"adler 1980)
in order to construct various types of dynamos, such as
$\alpha \Omega$ dynamo models.
Despite the fact that such models have been shown to be capable of producing a
large number of observationally relevant 
modes of behaviour, ranging from stationary to
chaotic (c.f. Brandenburg et al. 1989a,b;
Tavakol et al. 1995),
they nevertheless involve a number of unknown features such as the exact nature 
of the nonlinearities involved. Furthermore,
in order to clarify the origin of dynamical modes 
of behaviour observed in dynamo models, further simplifications
of these models have been considered, involving
low dimensional truncations of the governing PDE.
Such models have also been shown to be
capable of producing  a number of important features of 
stellar variability including periodic, intermittent and chaotic 
modes of behaviour (Zeldovich et al. 1983;
Weiss et al. 1984; Feudel et al. 1993).

Now given that these models are cheaper to integrate and 
more transparent to study, it would be very useful if we
could employ them as diagnostic tools in order to study the effects of 
introducing different parametrisations and nonlinearities involved.
The problem, however, is that 
these low dimensional models involve severe approximations,
and therefore in order to be able to take the results 
produced by them as physically relevant, it is important that they remain 
robust under changes which fall within the domain of the approximations assumed.
This is particularly of importance since on the basis of results
from dynamical systems theory, structurally stable systems 
are not everywhere dense in the space of dynamical systems
(Smale 1966), in the sense that small changes in models can
produce qualitatively important changes in their dynamics.
In this way the appropriate theoretical framework
for the construction of mathematical models and the analysis of
observational data may turn out to be that of structural 
fragility (Tavakol \& Ellis 1988; Coley \& Tavakol 1992; Tavakol et al. 1995). 

Here as examples of such changes we shall consider first changes
in the order of truncation and then changes in the details of the 
physics assumed. 
Regarding the former, a number of attempts have already been made 
to study the effects of increasing the truncation order on the 
resulting dynamics. For example, Schmalz \& Stix (1991) (hereafter
referred to as S\&S91) have looked at the detailed dynamics of 
the low dimensional truncations of the mean field
dynamo equations and have studied what happens as the order of the truncation
is increased, while Tobias et al. (1995) have employed
normal form theory to construct a robust minimal third order
model which exhibits both the modulation of basic cycles and chaos. These studies
have shown that low dimensional models can capture a number of important
dynamical features of the dynamo models.

Our aim in this paper is complementary to that of the above authors.
We take a detailed look at the results in S\&S91 and ask to what extent these results
remain robust as reasonable changes are made to the details of the physics
employed, and in each case we study how such changes affect the dynamical behaviour 
of different truncations. 
\section{Models with dynamical $\alpha$}
The starting point of the truncated dynamical $\alpha$ models
considered in S\&S91 is the mean field induction equation
\begin{equation}\label{induction}
\frac{\partial\vec{B}}{\partial t}=\nabla\,\times\,(\vec{v}\,\times\,\vec{B}+
\alpha\vec{B}-\eta_t\nabla\,\times\,\vec{B}),
\end{equation}
where $\vec{B}$ and $\vec{v}$ are the mean magnetic field and the mean
velocity, respectively. The turbulent magnetic diffusitivity $\eta_t$
and the coefficient $\alpha$, which relates the mean electrical current
arising in helical turbulence (the $\alpha$--effect) to the mean
magnetic field, both arise from the correlation of small scale
(turbulent) velocity and magnetic fields (Krause \& R\"adler 1980).

S\&S91 employ an axisymmetrical configuration with one spatial dimension $x$, 
which corresponds to a latitude coordinate and a longitudinal velocity
with a constant radial gradient (the vertical shear $\omega_0$).
The magnetic field takes the form
\begin{equation}\label{Bspherical}
\vec{B}=\left(0,B_{\phi},\frac{1}{R}\frac{\partial A_{\phi}}{\partial x}\right),
\end{equation}
where
$A_\phi$ is the $\phi$--component (latitudinal) of the magnetic vector
potential, $B_\phi$ the $\phi$--component of $\vec{B}$
and  $x$ is measured in terms of the stellar radius $R$.
These assumptions allow Eq. (\ref{induction}) to be split into
\begin{eqnarray}
\label{p1}
\frac{\partial A_{\phi}}{\partial t}&=&\frac{\eta_t}{R^2}
\frac{\partial^2A_{\phi}}{\partial x^2}+\alpha B_{\phi},\\
\label{p2}
\frac{\partial B_{\phi}}{\partial t}&=&\frac{\eta_t}{R^2}\frac{\partial^2 B_{\phi}}
{\partial x^2}+\frac{\omega_0}{R}\frac{\partial A_{\phi}}{\partial x}.
\end{eqnarray}
In S\&S91, $\alpha$ is divided into a static (kinematic) and a dynamic
(magnetic) part:
$\alpha=\alpha_0\cos x-\alpha_M(t)$, with its time-dependent part
$\alpha_M(t)$ satisfying an evolution equation in the form
\begin{equation}\label{dynamicalpha}
\frac{\partial \alpha_M}{\partial t}= {\Delta} (\alpha_M) + f(\vec{B}),
\end{equation}
where ${\Delta}$ is a damping operator and 
$f(\vec{B})$ is a pseudo-scalar that is quadratic 
in the magnetic filed.

It has been argued that the $\alpha$ effect is quenched by the current
helicity density $\vec{J}\cdot\vec{B}$,
which in turn is governed by a dynamical equation
(Kleeorin \& Ruzmaikin 1982; Zeldovich et al. 1983).
The reason the feedback (quenching) is not instantaneous is a consequence of the fact that the
magnetic helicity is conserved in the absence of diffusion or boundary
effects. Such models have been investigated recently by Kleeorin et al. (1995).
In S\&S91 a truncated version of
yet another model was studied, in which instead of the current helicity density, the
magnetic helicity density $\vec{A}\cdot\vec{B}$, or rather
$A_{\phi}B_{\phi}$, was used. Their model was motivated on heuristic grounds.
Bifurcation properties of a truncated version of a similar
model, but with a different damping term,
have been studied by Feudel et al. (1993).
Our present investigation is thus motivated partially by
the variety of models presented in the literature. It is important to
know what is the effect of the dynamical feedback and how the
different representations affect the results.

To proceed S\&S91 specify the feedback in the following way
\begin{equation}
f(\vec{B})\propto A_{\phi}B_{\phi}
\label{ab}
\end{equation}
and then look at various $N-$modal truncations of
these equations and study what happens to the dynamical
behaviour of the resulting systems as $N$ is increased.
 
To do this it is convenient to transform these equations into a
non-dimensional form. This can be done by employing  
a reference field $B_0$, measuring time
in units of $R^2/\eta_t$ and defining the following non-dimensional
quantities
\begin{eqnarray}
&&A=\frac{R\omega_0}{B_0\eta_t}A_{\phi},\quad B=\frac{B_{\phi}}{B_0},\quad
C=\frac{\alpha_M R^3\omega_0}{\eta_t^2},\\
&&\nu=\frac{\nu_t}{\eta_t},\quad
D=\frac{\alpha_0\omega_0 R^3}{\eta_t^2},\nonumber
\end{eqnarray}
where $\nu_t$ is the turbulent diffusivity.
Equations (\ref{p1}), (\ref{p2}) and
(\ref{dynamicalpha}) with the damping operator taken to be
\begin{equation}
{\Delta} = \frac{\nu_t}{R^2}
\frac{\partial^2 }{\partial x^2},
\end{equation}
can then be rewritten in the following
non-dimensional forms:
\begin{eqnarray}
\label{1}
\frac{\partial A}{\partial t}&=&\frac{\partial^2 A}
{\partial x^2}+DB\cos x - CB,\\
\frac{\partial B}{\partial t}&=&\frac{\partial^2 B}{\partial x^2}+
\frac{\partial A}{\partial x},\\
\label{3}
\frac{\partial C}{\partial t}&=&\nu\frac{\partial^2 C}{\partial x^2}+
AB.
\end{eqnarray}

Now considering the interval $0\le x\le\pi$
(which corresponds to the full range of latitudes), taking the
boundary conditions at $x=0$ and $x=\pi$ to be given by $A=B=C=0$ and
using a spectral expansion of the form
\begin{eqnarray}
A=\sum_{n=1}^{N}A_n(t)\sin nx,\\
B=\sum_{n=1}^{N}B_n(t)\sin nx,\\
C=\sum_{n=1}^{N}C_n(t)\sin nx,
\end{eqnarray}
allows the set of Eqs. (\ref{1}--\ref{3}) to be transformed into
the form
\begin{eqnarray}
\label{main3_1}
\frac{\partial A_n}{\partial t}&=&-n^2A_n+\frac{D}{2}(B_{n-
1}+B_{n+1})\\&&+\sum_{m=1}^{N}\sum_{l=1}^{N}F(n,m,l)B_mC_l,\nonumber\\
\label{main3_2}
\frac{\partial B_n}{\partial t}&=&-n^2B_n+\sum_{m=1}^{N}G(n,m)A_m,\\
\label{main3_3}
\frac{\partial C_n}{\partial t}&=&-\nu n^2 C_n
-\sum_{m=1}^{N}\sum_{l=1}^{N}F(n,m,l)A_mB_l,
\end{eqnarray}
where
\begin{eqnarray}
&&F(n,m,l)=\\&&\frac{8nml}{\pi(n+m+l)(n+m-l)(n-m+l)(n-m-l)}\nonumber,
\end{eqnarray}
if $n+m+l$ is odd and $F(n,m,l)=0$ otherwise and
\begin{equation}
G(n,m)=\frac{4nm}{\pi(n^2-m^2)},
\end{equation}
if $n+m$ is odd and $G(n,m)=0$ otherwise.

These rules enable the system to describe fields which are strictly
symmetric (i.e. having only components $B_n$ with odd $n$
and $A_n$ and $C_n$ with even $n$)
or strictly antisymmetric (i.e. having only components 
$A_n$ with odd $n$ and $B_n$ and $C_n$ with
even $n$) with respect to $x=\pi/2$, provided the initial conditions have either of these
parities.

Using these equations, S\&S91 studied a number of such truncations
numerically by  varying the dynamo number $D$ at each truncation $N$.
Their main conclusions were:

\begin{enumerate}
      \item With the choice of the driving term $f$ given by Eq. (\ref{ab})
            the antisymmetric truncation with the
            smallest non-trivial indices is identical with 
            the Lorenz system (Lorenz 1963).

      \item Different truncations are capable of producing 
            stationary, oscillatory and chaotic modes of behaviour.
            They also make observations about the changes in the route to
            chaos, and conclude that, as $N$ is increased, the 
            route changes from period doubling to the
            Ruelle--Takens--Newhouse 
            scenario (Ruelle \& Takens 1971; Newhouse et al. 1978).
      
      \item The qualitative behaviour of the truncations 
            stabilises as the number of
            modes is increased and in particular for $N>6$.
            As an example they observe that as $N$ is increased
            the limit cycles remain stable for larger dynamo numbers.

      \item They also discuss very briefly the $D<0$ case, observing 
            that the $N=2$ case is always a stable fixed point
            and that for $N \ge 6$ the antisymmetric limit cycle
            becomes unstable via a saddle node
            bifurcation\footnote{Care must be taken when speaking of
            antisymmetric solutions. In our studies we mean 
            strictly antisymmetric solutions,
            while in S\&S91 these also refers to the antisymmetric part of mixed
            parity solutions.}.
            
\end{enumerate}

Now, as mentioned above, there are arguments 
in support of both the form of the driving term
as well as the damping term being different
(Kleeorin et al. 1995).
So as a first step, we shall study, in the next section, how robust the
results in S\&S91 are with respect to various 
physically justified changes in the driving term that have
been considered in the literature in
Eq. (\ref{dynamicalpha}).
In Sect. \ref{T} we study the effects of changes in the 
damping term.

\section{Robustness with respect to changes in the driving term}

The general physically motivated choice for the driving term
is given by Kleeorin \& Ruzmaikin (1982),
Zeldovich et al. (1983) and Kleeorin et al. (1995)
to be in the form 
\begin{equation}
f= W_1 \vec{J}\cdot\vec{B} + W_2 \alpha {|\vec{B}|}^2 ,
\end{equation}
where $W_1$ and $W_2$ are constants.
To study the effects of each term separately, we
shall proceed by considering the cases $W_1\ne 0$ ($W_2 = 0$) and  
$W_1\ne 0$ ($W_2\ne 0$) in the following sections.
\subsection{Case (I): $f= W_1 \vec{J}\cdot\vec{B} + W_2 \alpha {|\vec{B}|}^2$,
with $W_1\ne 0$ $(W_2=0)$}
\label{caseI}

Taking $f$ to be of the form $f\propto 
\vec{J}\cdot\vec{B}$, substituting for
$\vec{B}$ from Eq. (\ref{Bspherical}) and recalling that 
$\vec{J}=\nabla\times\vec{B}$ we obtain 
\begin{equation}
\vec{J}\cdot\vec{B}=\left(\nabla\times\vec{B}\right)\cdot\vec{B}
=\frac{B_0^2\eta_t}{R^3\omega_0}
\left(\frac{\partial A}{\partial x}\frac{\partial B}{\partial x}
-\frac{\partial^2A}{\partial x^2}B\right),
\end{equation}
which allows Eq. (\ref{3}) to be written as
\begin{equation}
\label{jB}
\frac{\partial C}{\partial t}=\nu\frac{\partial^2 C}
{\partial x^2}+\frac{\partial A}{\partial x}\frac{\partial B}
{\partial x}-\frac{\partial^2A}{\partial x^2}B.
\end{equation}

Proceeding in a similar way as in previous section
we obtain an identical set of differential equations to those
obtained in S\&S91, except that Eq. (\ref{main3_3}) is
now changed to
\begin{equation}
\frac{\partial C_n}{\partial t}=-\nu n^2 C_n
-\sum_{m=1}^{N}\sum_{l=1}^{N}H(n,m,l)A_mB_l,
\label{Cn}
\end{equation}
where
\begin{eqnarray}
\label{H}
&&H(n,m,l)=
\\&&\frac{4}{\pi} {\frac {{n} {m} {l}  (- {n}^{2} + 3 {m}
^{2} + {l}^{2} )}{( {n} + {l} + {m} ) ( {n} + {l} - {m} ) ( {
n} - {l} + {m} ) ( {n} - {l} - {m} )}}\nonumber,
\end{eqnarray}
if $n+m+l$ is odd and $H(n,m,l)=0$ otherwise.
The function $H$ is clearly different from $F$ unless 
$F=0$, in which case $H$ is also equal to zero. 

For this system we
can study also the pure 
antisymmetric and symmetric solutions,
but for the sake of comparison with the results 
in S\&S91 we confined ourselves to the 
antisymmetric solutions. 

Now for the case of $N=2$, the Eqs.
(\ref{main3_1}), (\ref{main3_2}) and (\ref{Cn}) become
\begin{eqnarray}
\frac{dA_1}{dt}&=&-A_1+\frac{DB_2}{2}-\frac{32 B_2C_2}{15\pi},\\
\frac{dB_2}{dt}&=&-4 B_2+\frac{8 A_1}{3\pi},\\
\frac{dC_2}{dt}&=&-4\nu C_2+\frac{16 A_1B_2}{5\pi},
\end{eqnarray}
which upon using the transformations
\begin{equation}
A_1={\frac {15\,\sqrt {6}{\pi }^{2}}{64}}Y,\quad
B_2={\frac {5\,\sqrt {6}\pi }{32}}X,\quad
C_2={\frac {45\,{\pi }^{2}}{64}}Z,
\end{equation}
result, as in S\&S91, in the usual Lorenz equations
(Lorenz 1963), with the control parameters given by
$\sigma=4$, $b=4\nu$, and $r=D/3\pi$. To be compatible with S\&S91
we also used $\nu=0.5$ 
throughout\footnote{These authors seem to confine themselves
to this value of $\nu$ in order to obtain chaotic behaviour,
for which one requires $\sigma>b+1$ (Sparrow 1982). This amounts
to the expectation that $\alpha$ relaxes much more slowly than the magnetic
field.}.

Since our aim is to study the qualitative effects brought
about by the
changes in the form of $f$, we will not delve deeply into the details 
of the dynamics, such as the routes to chaos, and concentrate
instead on the occurrence of equilibrium, periodic (including
quasiperiodic) and chaotic regimes. Accordingly, the tools
we employ are the time series and the 
spectra of Lyapunov exponents. The latter
is particularly useful as a relatively sensitive tool
to characterise the dynamics, with the Lyapunov spectra of the 
types $(-,-,-,\ldots)$, $(0,-,-,\ldots)$,
$(0,0,-,\ldots)$ and $(+,0,-,\ldots)$
corresponding to equilibrium, periodic, quasiperiodic (with
two periods) and chaotic regimes respectively.
Also to keep the numerical costs reasonable,
the resolution of $D$ in all the figures was,
unless stated otherwise, taken to be $D=5$.

Now given the fact that in many astrophysical settings
(including that of the sun) the sign of the dynamo number
is not known, we shall also study the 
effects of changes in the sign of $D$.

We note also that the $\alpha \Omega$ dynamo concept becomes invalid if $D$
exceeds a certain limit (Choudhuri 1990). Furthermore, in general,
as $D$ is increased more modes (higher $N$) are required to
achieve convergence (numerically bounded solutions).

\subsubsection{Results for positive dynamo numbers}

\begin{figure}
\centerline{\def\epsfsize#1#2{0.33#1}\epsffile[90 73 742 558]{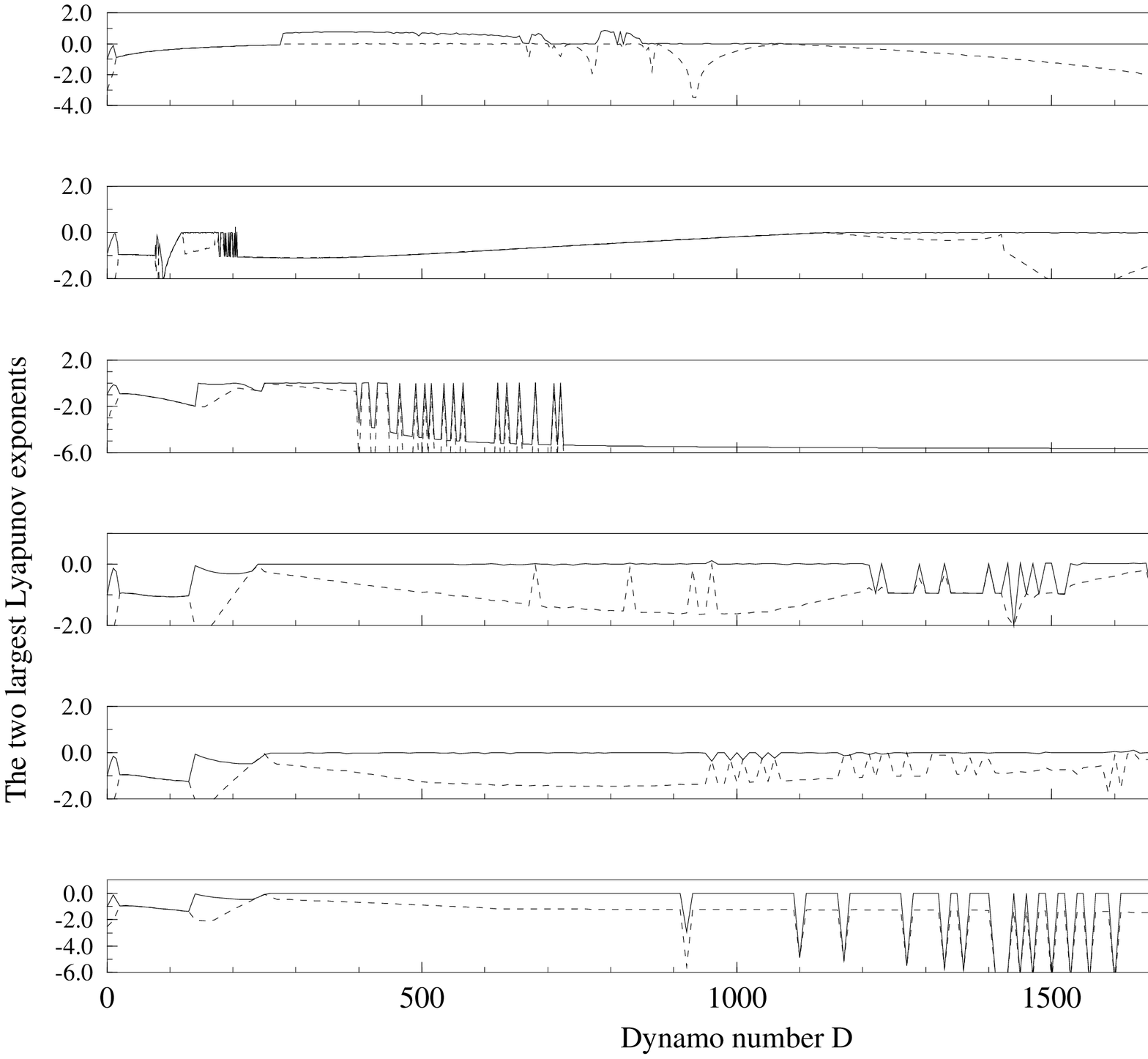}}
\caption[]{\label{jb_plus}
Graphs of the two largest Lyapunov exponents for $N=2$, 4, 6, 7, 8 and
10 (increasing downwards) for the case where $f\propto \vec{J}\cdot\vec{B}$
and $D>0$}
\end{figure}

For the sake of comparison 
with S\&S91, we studied the dynamics
of the system (\ref{main3_1}, \ref{main3_2}, \ref{Cn}),
for different values of the truncation
order $N$. A summary of our numerical results 
is given in Fig. \ref{jb_plus} which is a plot of the two largest
Lyapunov exponents as a function of the dynamo number
for different truncations. In the following figures, the largest
Lyapunov exponent is depicted by a solid line and its negative,
zero and positive values indicate equilibrium, periodic and chaotic
regimes. The simultaneous vanishing of the second Lyapunov exponent
would imply the presence of quasiperiodic motion with
two frequencies (i.e. motion on a 2--torus). It was not necessary to plot the
third exponent, since no motion on $T^3$ or higher dimensional tori was
observed which is not surprising in view of the 
results of Newhouse et al. (1978).

\begin{figure}
\centerline{\def\epsfsize#1#2{0.33#1}\epsffile[90 55 742 556]{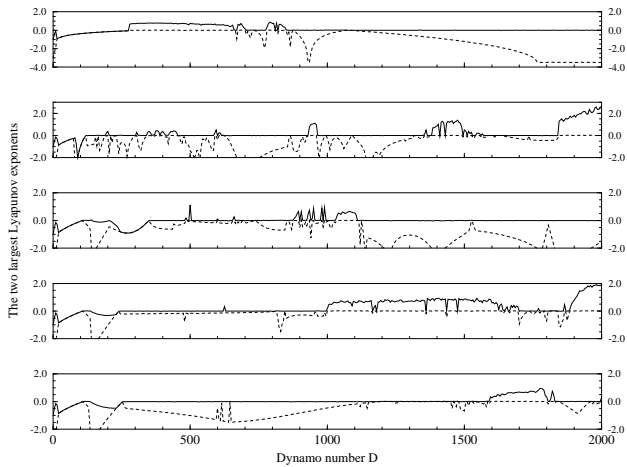}}
\caption[]{\label{ab_plus}
Graphs of the two largest Lyapunov exponents for $N=2$, 4, 6, 7, and
8 (increasing downwards) for the case where $f\propto A_{\phi}B_{\phi}$
and $D>0$}
\end{figure}

For a more transparent comparison,
we have also produced in Fig. \ref{ab_plus} an analogous figure
for the system considered in S\&S91.
As the comparison of the Figs. \ref{jb_plus} and \ref{ab_plus}
shows, the main differences produced by the replacement of $A_{\phi}B_{\phi}$
by $\vec{J}\cdot\vec{B}$ are as follows:

\begin{enumerate}
\item The chaotic regimes become less likely in the $\vec{J}\cdot\vec{B}$
case, in the sense that
the intervals of the dynamo number $D$ over which the system is
chaotic decrease dramatically. 
\item
There exist indications for the presence of ``multiple attractors''
over substantial intervals of $D$,
consisting of equilibrium and periodic states.
These can be seen as regions of spiky behaviour in the solid line 
in Fig. \ref{jb_plus}, for certain truncations ($N =4,6,7,8,10$). The
behaviour of the system alternates between fixed 
point solutions (where all exponents are
negative) and periodic orbits (where only the first one is zero) as the
dynamo number $D$ is slightly changed.

The presence of such behaviour is potentially of great interest since
it suggests that there exist intervals of $D$ in which small changes 
in $D$ can drastically change the behaviour of the system. This is
also interesting, if one considers settings in which $D$
or the initial conditions (IC) can vary slightly, but randomly, as 
the resulting behaviour would look
very much like intermittency. To highlight this we have plotted in
Fig. \ref{fragility} the behaviour of the $N=4$ 
truncation as a function of small changes
in the dynamo number and the IC. 
As can be seen, small changes in either $D$ or IC can produce 
important changes in the behaviour of the system.
This therefore shows that there are substantial regions of 
$D$ over which the behaviour of the system is sensitive to
small changes in $D$ and IC. Further, we have checked that
this fragility is itself 
robust in the sense that taking a finer mesh of $D$ 
does not qualitatively change this overall behaviour.

\begin{figure}
\centerline{\def\epsfsize#1#2{0.33#1}\epsffile[0 0 612 472]{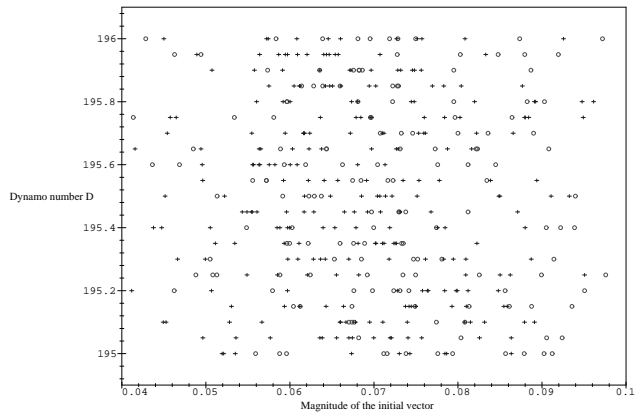}}
\caption[]{\label{fragility}
Fragility in the dynamics with respect to small changes in 
the dynamo number $D$ and
the magnitude of the initial vector $(A_n, B_n, C_n)$. A cross represents a fixed
point while a circle represents a limit cycle. The initial conditions
correspond to randomly chosen vectors of specified magnitude}
\end{figure}

\item
Regarding the overall behaviour of the systems with respect to 
increases in $N$, we observe the following.
For small dynamo numbers, the behaviour seems to settle down to
equilibrium and periodic states as $N$ is increased.
For example as can be seen from Fig. \ref{jb_plus},
for dynamo numbers up to $D \approx 900$, the behaviour
settles down for $N \ge 7$. For larger values of $D$, however,
we observe an increase in the dominance of the ``multiple 
attractor'' regime for the values of $N$ considered here.
It is likely, however, that with increasing $N$, these intervals only
establish themselves at higher values of $D$.
\item
The transition to chaos appears to be very abrupt in the 
$N=2$ case, with the system going from a fixed point into
a chaotic regime very rapidly, at least to within a 
resolution of  $\Delta D \approx 10^{-4}$,
with no intermediate behaviour being observed.
For the case $N=3$ the system goes from a fixed
point $\to$ limit cycle $\to$ chaos.
For still higher $N$, our calculations indicate that
chaos becomes scarce.
\item
Chaotic regions were also found in the ``multiple attractors'' region, which 
were fragile with respect to small changes in the IC and
the choice of $D$. 
\end{enumerate}

\subsubsection{Results for negative dynamo numbers}

\begin{figure}
\centerline{\def\epsfsize#1#2{0.33#1}\epsffile[90 73 742 553]{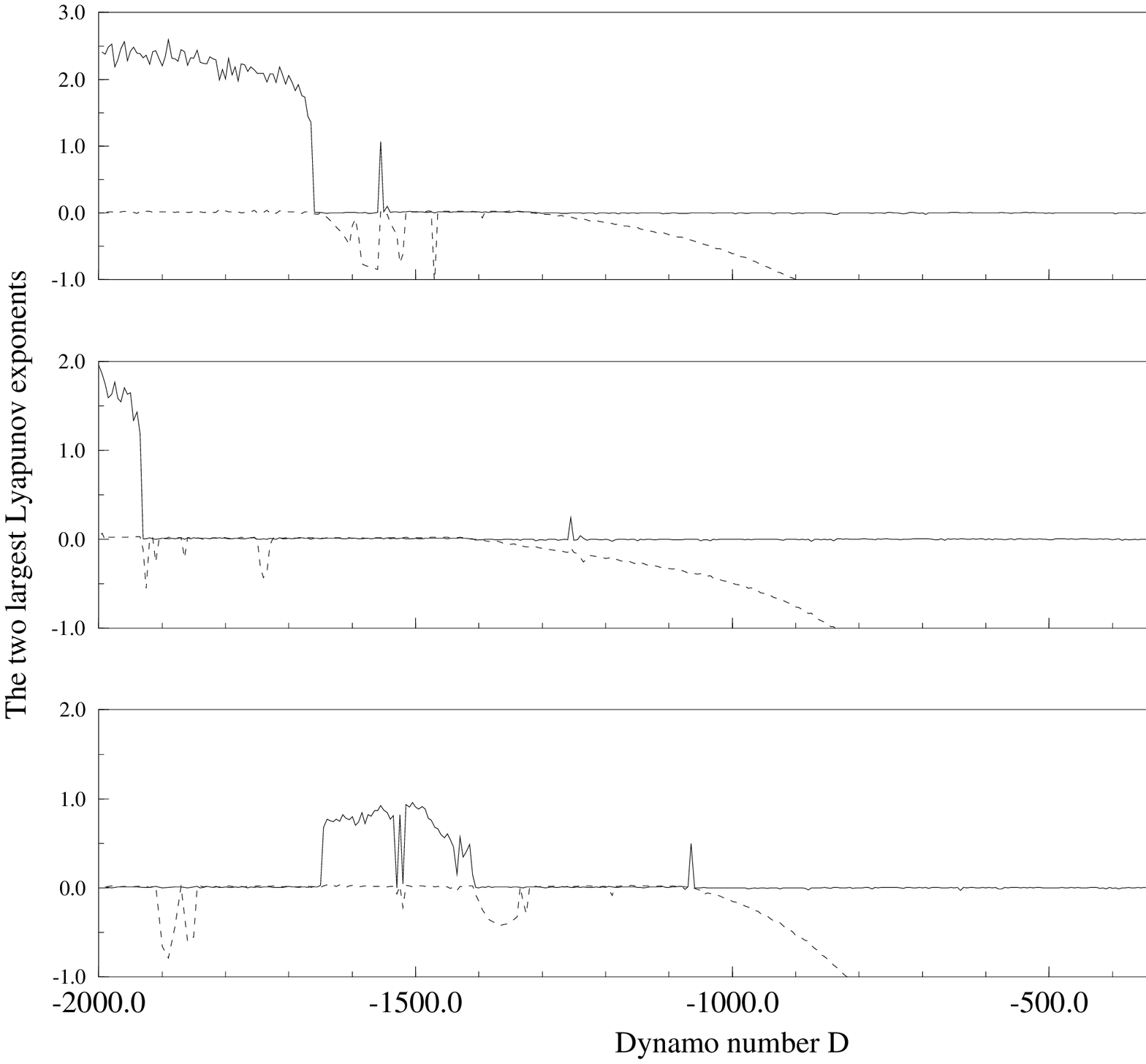}}
\caption[]{\label{jb_minus}
Graphs of the two largest Lyapunov exponents for $N=7$, 8, and
10 (increasing downwards) 
for the case when $f\propto \vec{J}\cdot\vec{B}$ and $D<0$}
\end{figure}

\begin{figure}
\centerline{\def\epsfsize#1#2{0.33#1}\epsffile[90 73 742 564]{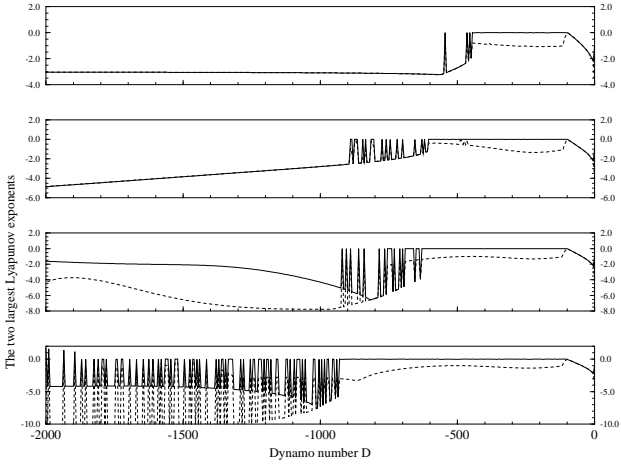}}
\caption[]{\label{ab_minus}
Graphs of the two largest Lyapunov exponents for $N=4$, 6, 7, and
8 (increasing downwards) for the case when $f\propto A_{\phi}B_{\phi}$ and $D<0$}
\end{figure}

Our results for the negative dynamo numbers are shown in Fig. \ref{jb_minus}.
Also, in view of the sparseness of the results reported in 
S\&S91 for the models with negative dynamo numbers,
we present Fig. \ref{ab_minus}
as an analogous figure for their case.

The main features of these models are:
\begin{enumerate}
\item
The chaotic regimes seem to become less likely in the 
$A_{\phi}B_{\phi}$ case. In fact, for the mesh size in $D$ taken here,
we only observed chaotic solutions in the case of
$N=8$ and then only for very high dynamo numbers.
\item
There are substantial intervals (in $D$) of
``multiple attractors'' (consisting of equilibrium and periodic states)
for the $A_{\phi}B_{\phi}$ case.
\item
In both cases the behaviour for $D\ga -900$ stabilises
as $N$ is increased. This occurs for $N\ge 2$ for the equilibrium regime
and $N\ge 7$ for periodic regime.
These results also indicate that there 
are parallels between the $A_{\phi}B_{\phi}$ case with negative dynamo numbers
and the $\vec{J}\cdot\vec{B}$ case with positive 
dynamo numbers. In both cases, multiple attractor regions
seem to dominate for large $D$ values, as $N$ is increased.
\item
For high $N (\ge 7)$ in the $\vec{J}\cdot\vec{B}$ case,
the transition is from $T^2$ to chaotic behaviour. 
This does not seem not true for
$N=4, 5, 6$ where the chaotic behaviour seems to appear abruptly.
\end{enumerate}

\subsection{Case (II): $f= W_1 \vec{J}\cdot\vec{B} + W_2 \alpha |{\vec{B}}|^2$,
with $W_1\ne 0$ and $W_2\ne 0$}
\label{caseII}
To study the effects of including the $\alpha|\vec{B}|^2$
term, we use the dynamic $\alpha$ equation from 
Kleeorin \& Ruzmaikin (1982) 
without a damping term\footnote{The inclusion of this term will
be studied in Sect. \ref{T}.} proportional to $1/T$

\begin{eqnarray}
\frac{\partial \alpha_M}{\partial t}&=&
\frac{\nu_t}{R^2}\frac{\partial^2 \alpha_M}{\partial x^2}\\
&&-\frac{1}{\mu_0\rho}\left(\left(\nabla\times\vec{B}\right)\cdot\vec{B}
-\frac{\alpha_0\cos x - \alpha_M}{\beta}|\vec{B}|^2\right),
\nonumber
\end{eqnarray}
where $\beta$ is the combined (turbulent plus ohmic) diffusion of the
field, $\rho$ the density of the medium and $\mu_0$ the magnetic constant. 

Now using expression (\ref{Bspherical}) for $\vec B$
and turning the system in a non-dimensional form
using the same transformations as before, we obtain
\begin{eqnarray} 
&&\frac{\alpha_0\cos x-\alpha_M}{\beta}|\vec{B}|^2=\\
&&\frac{B_0^2\eta_t^2}{R^3\omega_0\beta}(D\cos x-C)\left(B^2+\Gamma_1
\left(\frac{\partial A}{\partial x}\right)^2\right)\nonumber,
\end{eqnarray}
where $\Gamma_1=\frac{\eta_t^2}{R^4\omega_0^2}$ is a dimensionless
constant. This allows the analogue of the Eq. (\ref{3}) to be written in the form
\begin{eqnarray}\label{alphab2}
\frac{\partial C}{\partial t}=\nu\frac{\partial^2 C}{\partial x^2}-
\Gamma_2\times \left(\frac{\partial A}{\partial x}\frac{\partial B}{\partial x}
-\frac{\partial^2A}{\partial x^2}B\right)\\
+\Gamma_3\times(D\cos x-C)\left(B^2+
\Gamma_1\left(\frac{\partial A}{\partial x}\right)^2\right)\nonumber,
\end{eqnarray}
where $\Gamma_2=\frac{R^2B_0^2}{\eta_t^2\mu_0\rho}$ and
$\Gamma_3=\frac{R^2B_0^2}{\beta\eta_t\mu_0\rho}$ are dimensionless
constants.
Considering the same boundary conditions and spectral expansions as in
the $\vec{J}\cdot\vec{B}$ case, Eq. (\ref{alphab2}) 
becomes
\begin{eqnarray}\label{Cn_2}
\frac{\partial C_n}{\partial t}&=&-\nu n^2 C_n
+\Gamma_2\left(\sum_{m=1}^{N}\sum_{l=1}^{N}
H(n,m,l) A_m B_l\right)\\
&&+\Gamma_3\sum_{m=1}^N\sum_{l=1}^N D B_m B_l H_1(n,m,l)\nonumber\\
&&+\Gamma_3\Gamma_1 \sum_{m=1}^N\sum_{l=1}^N m l D A_m A_l H_2(n,m,l)\nonumber\\
&&-\Gamma_3\sum_{m=1}^N\sum_{l=1}^N\sum_{k=1}^N C_m B_l B_k H_3(n,m,l,k)\nonumber\\
&&-\Gamma_3\Gamma_1\sum_{m=1}^N\sum_{l=1}^N\sum_{k=1}^N l k C_m A_l A_k H_4(n,m,l,k)\nonumber,
\end{eqnarray}
where $H$'s are given by
\begin{eqnarray}
H_1(n,m,l)&=&\frac{1}{\pi}
\left(\frac{m+1}{(m+1+l-n)(m+1-l+n)}\right.\\
&&-\frac{m+1}{(m+1+l+n)(m+1-l-n)}\nonumber\\
&&+\frac{m-1}{(m-1+l-n)(m-1-l+n)}\nonumber\\
&&\left.-\frac{m-1}{(m-1+l+n)(m-1-l-n)}\right),\nonumber\\
H_2(n,m,l)&=&\frac{1}{\pi}
\left(\frac{n-l}{(n-l+m+1)(n-l-m-1)}\right.\\
&&+\frac{n-l}{(n-l+m-1)(n-l-m+1)}\nonumber\\
&&+\frac{n+l}{(n+l+m+1)(n+l-m-1)}\nonumber\\
&&+\left.\frac{n+l}{(n+l+m-1)(n+l-m+1)}\right),\nonumber\\
H_3(n,m,l,k)&=&\frac{1}{4}
\left[\delta(m-l,k-n)-\delta(m-l,k+n)\right.\\
&&\left.-\delta(m+l,k-n)+\delta(m+l,k+n)\right],\nonumber\\
H_4(n,m,l,k)&=& \frac{1}{4}
\left[\delta(m-n,l-k)+\delta(m-n,l+k)\right.\\
&&\left.-\delta(m+n,l-k)-\delta(m+n,l+k)\right].\nonumber
\end{eqnarray}

Note that $\delta(n,m)$ is 1 if $n-m=0$ but 2 if $n=m=0$ and $H_1=H_2=0$
if $n+m+l+1$ is even.

\subsubsection{Results}

Our results of the study of the system (\ref{main3_1}), (\ref{main3_2}) and (\ref{Cn_2}) 
for positive dynamo numbers are depicted in Table \ref{tablealpha1}.
As can be seen, the effect of the inclusion of the $\alpha|\vec{B}|^2$ term
is dramatic and seems to eliminate the possibility of chaotic behaviour
for all $N$.

\begin{table}
\caption[]{\label{tablealpha1}
Results for the case (II) for $D>0$. $D_1$ indicates where the origin becomes
unstable as a fixed point and $D_2$ the dynamo number where all fixed points become
unstable and the solution becomes periodic}
\begin{flushleft}
\begin{tabular}{ccc}
\hline\noalign{\smallskip}
$N$ & $D_1$ & $D_2$ \\
\noalign{\smallskip}
\hline\noalign{\smallskip}
2 & $ 10$ & $>2000$ \\
3 & $ 15$ & $>2000$ \\
4 & $115$ & $115$ \\
5 & $205$ & $205$ \\
6 & $205$ & $240$ \\
7 & $235$ & $240$ \\
8 & $250$ & $250$ \\
\hline\noalign{\smallskip}
\end{tabular}
\end{flushleft}
\end{table}

For the lower truncations of $N=2$ and $3$, we only observe 
fixed point solutions
for all $D$ up to $D\approx 2000$. For higher order truncations, with
moderate $D$, there is a sequence of fixed
points followed by stable periodic cycles.

The corresponding results for the negative dynamo numbers are shown
in the Table \ref{tablealpha2}, and again this is very
similar to Table \ref{tablealpha1} with no evidence
for chaotic behaviour at small and moderate $D$. 
In this case the $N=2$ system has the origin as the fixed point
for $D$ down to $-2000$.

\begin{table}
\caption[]{\label{tablealpha2}
Results for the case (II) for $D<0$. $D_1$ indicates where the origin becomes
unstable as a fixed point and $D_2$ the dynamo number where all fixed points become
unstable and the solution becomes periodic}
\begin{flushleft}
\begin{tabular}{ccc}
\hline\noalign{\smallskip}
$N$ & $D_1$ & $D_2$ \\
\noalign{\smallskip}
\hline\noalign{\smallskip}
2 & $<-2000$ & $<-2000$ \\
3 & $-70$ & $-80$ \\
4 & $-85$ & $-95$ \\
5 & $-85$ & $-95$ \\
6 & $\approx -95$ & $\approx -100$ \\
7 & $\approx -95$ & $\approx -100$ \\
8 & $\approx -95$ & $\approx -100$ \\
\hline\noalign{\smallskip}
\end{tabular}
\end{flushleft}
\end{table}

\section{Case (III): Robustness with respect to changes in the damping term}
\label{T}

In this section we employ the equation
proposed by Kleeorin et al. (1995) in the form 

\begin{eqnarray}
\label{alphaT}
\frac{\partial \alpha_M}{\partial t}&=&-\frac{\alpha_M}{T}\\
&&-\frac{1}{\mu_0\rho}\left(\left(\nabla\times\vec{B}\right)\cdot\vec{B}
-\frac{\alpha_0\cos x - \alpha_M}{\beta}|\vec{B}|^2\right),
\nonumber
\end{eqnarray}
as the evolutionary equation for the back reaction of the magnetic field
on the time dependent part of $\alpha$. In the above equation $T$ is the
characteristic time on which the small scale magnetic helicity changes,
which is typically much longer than the turbulent diffusion time scale.

Using the same expression for $\vec{B}$ from Eq. (\ref{Bspherical}) and
proceeding in the same way as in the previous cases we obtain the 
differential equations for $C_n$
to be

\begin{eqnarray}\label{Cn_T}
\frac{\partial C_n}{\partial t}&=&-\Gamma_4 C_n
+\Gamma_2\left(\sum_{m=1}^{N}\sum_{l=1}^{N}
H(n,m,l) A_m B_l\right)\\
&&+\Gamma_3\sum_{m=1}^N\sum_{l=1}^N D B_m B_l H_1(n,m,l)\nonumber\\
&&+\Gamma_3\Gamma_1 \sum_{m=1}^N\sum_{l=1}^N m l D A_m A_l H_2(n,m,l)\nonumber\\
&&-\Gamma_3\sum_{m=1}^N\sum_{l=1}^N\sum_{k=1}^N C_m B_l B_k H_3(n,m,l,k)\nonumber\\
&&-\Gamma_3\Gamma_1\sum_{m=1}^N\sum_{l=1}^N\sum_{k=1}^N l k C_m A_l A_k H_4(n,m,l,k)\nonumber,
\end{eqnarray}
where $\Gamma_4=\frac{R^2}{\eta_t T}$ is a dimensionless constant.

\subsection{Results}

Our results of the study of the system
(\ref{main3_1}), (\ref{main3_2}) and (\ref{Cn_T})
are shown in Tables \ref{tableT1} and \ref{tableT2}.
Although more modes are required in order to
obtain convergence for higher dynamo numbers,
the results shown in Table \ref{tableT1} and \ref{tableT2} seem 
to indicate that this type of change in the damping
term does not produce qualitative changes in the
behaviour of the system.
This is reasonable, since the functional forms of the 
modal equations are quite similar in Eqs. (\ref{Cn_2}) and (\ref{Cn_T}).

The inclusion of the $\alpha_M/T$ term does not
change the qualitative behaviour of
the smaller truncations ($N=2$ and $3$ for $D>0$
and $N=2$ for $D<0$, where we observe only fixed points
as before). At moderate dynamo numbers, the qualitative behaviour 
is almost the same and remains periodic for $|D|>|D_2|$, but $D_2$
is changed slightly.

\begin{table}
\caption[]{\label{tableT1}
Results for the case (III) for $D>0$. $D_1$ indicates where the origin becomes
unstable as a fixed point and $D_2$ the dynamo number where all fixed points become
unstable and the solution is a periodic orbit}
\begin{flushleft}
\begin{tabular}{ccc}
\hline\noalign{\smallskip}
$N$ & $D_1$ & $D_2$ \\
\noalign{\smallskip}
\hline\noalign{\smallskip}
2 & 10 & $>2000$ \\
3 & 15 & $>2000$ \\
4 & $110$ & $115$ \\
5 & $170$ & $170$ \\
6 & $\approx 200$ & $\approx 200$ \\
7 & $\approx 200$ & $\approx 200$ \\
8 & $\approx 200$ & $\approx 200$ \\
\hline\noalign{\smallskip}
\end{tabular}
\end{flushleft}
\end{table}

\begin{table}
\caption{\label{tableT2}
Results for the case (III) for $D<0$. $D_1$ indicates where the origin becomes
unstable as a fixed point and $D_2$ the dynamo number where all fixed points become
unstable and the solution is a periodic orbit}
\begin{flushleft}
\begin{tabular}{ccc}
\hline\noalign{\smallskip}
$N$ & $D_1$ & $D_2$ \\
\noalign{\smallskip}
\hline\noalign{\smallskip}
2 & $<2000$ & $<2000$ \\
3 & $-70$ & $-80$ \\
4 & $-80$ & $-95$ \\
5 & $-95$ & $-95$ \\
6 & $\approx -95$ & $\approx -105$ \\
7 & $\approx -95$ & $\approx -110$ \\
8 & $\approx -95$ & $\approx -110$ \\
\hline\noalign{\smallskip}
\end{tabular}
\end{flushleft}
\end{table}

We also note that all systems considered here, in particular Cases (II) and (III), have a
common pattern of behaviour, namely that as $D$ is increased, $A$ and $B$
oscillate with slowly increasing amplitudes about zero.
On the other hand, $\alpha_M$ oscillates with an increasing amplitude
around a rapidly increasing average. Also if $D>0$, $\alpha_M$
oscillates about a positive average and about a negative average for $D<0$.

\section{Conclusions}

We have studied the robustness of truncated $\alpha \Omega$ dynamos 
including a dynamic $\alpha$ equation, with respect to physically motivated
changes in the driving term and a change in the 
damping term appearing in the dynamical $\alpha$ equation. We studied these
systems with respect to changes in the dynamo number $D$,
the truncation order $N$ and the IC.
Our results show that the changes in the driving term have
important effects on the dynamical behaviour of the
resulting systems.
In particular we find that 

\begin{itemize}
\item chaos is much less
likely in systems with a driving term
of the form $\vec{J}\cdot\vec{B}$ (with positive $D$),
as opposed to those involving $A_{\phi}B_{\phi}$. 
\item the inclusion of the
$\alpha|\vec{B}|^2$ term has a dramatic effect in that it 
suppresses the possibility
of chaotic behaviour at moderate dynamo numbers.
\item  
changes in the sign of the dynamo number can also produce 
important changes. In the case where the driving term
is given by $A_{\phi}B_{\phi}$, using $D<0$
makes chaotic behaviour much less likely (which seems
to be the mirror image of the case where the driving term
given by $\vec{J}\cdot\vec{B}$ and $D>0$).
\item in case (I) there exists substantial intervals of
$D$ for which the systems seem to possess "multiple attractors"
(consisting of equilibrium and periodic states). As a 
result small changes in either $D$ or the IC can produce
important changes in these regimes.
This form of fragility can be of importance, especially in presence of 
noise, where the system would behave in an intermittent way.
\end{itemize}

Finally to recapitulate our motivation for studying
different formulations of 
dynamic $\alpha$ feedback, we note that even the usual 
expression for the driving term, $\vec{J}\cdot\vec{B}$, derived from first principles
could still be inappropriate, as it involves  uncontrolled
approximations. 
However, it is clear that $f$ has to be a pseudo-scalar (because $\alpha$ is a
pseudo-scalar), and the most obvious possibilities are indeed the ones
that we have studied.
Our investigations have shown that the actual choice can
significantly alter the overall conclusion. Therefore, all conclusions,
especially those concerning the occurrence of chaos, should be taken with
utmost care.

\begin{acknowledgements} EC is supported by grant  BD / 5708 / 95 -- 
Program PRAXIS XXI, from JNICT -- Portugal.
RT benefited from SERC UK Grant No. H09454. This research also benefited
from the EC Human Capital and Mobility (Networks) grant ``Late type stars:
activity, magnetism, turbulence'' No. ERBCHRXCT940483.
\end{acknowledgements}



\end{document}